\documentstyle[pra,aps,twocolumn]{revtex}
\begin{document}
\title{Quantum State Separation, Unambiguous Discrimination and Exact Cloning}
\author{Anthony Chefles\thanks{e-mail: tony@phys.strath.ac.uk} and Stephen M. Barnett}
\address{Department of Physics and Applied Physics, \\ University of Strathclyde,
       Glasgow G4 0NG, Scotland}
\input epsf
\epsfverbosetrue

\def\id{\hat{\leavevmode\hbox{\small1\kern-3.2pt\normalsize1}}}%

\maketitle

\begin{abstract}
Unambiguous discrimination and exact cloning reduce the square-overlap between quantum states, exemplifying the more general type of procedure we term {\em state separation}.  We obtain the maximum probability with which two equiprobable quantum states can be separated by an arbitrary degree, and find that the established bounds on the success probabilities for discrimination and cloning are special cases of this general bound.  The latter also gives the maximum probability of successfully producing $N$ exact copies of a quantum system whose state is chosen secretly from a known pair, given $M$ initial realisations of the state, where $N{\ge}M$.  We also discuss the relationship between this bound and that on unambiguous state discrimination. 

\end{abstract}

\mbox{}

One of the most characteristic features of quantum information is the fact that the amount of information required to specify the state of a quantum system is far greater, in fact infinitely greater, than the amount of information that can actually be extracted by measurement\cite{Caves}.  In other words, it is impossible to experimentally determine the state of a quantum system, if we do not know an orthonormal basis to which it belongs.  This implies that an infallible procedure for copying the state cannot exist either\cite{WZurek}.  If  perfect cloning were possible, then iteration of the cloning procedure would yield an arbitrarily large ensemble of independent systems prepared in the same state, which could be inferred through the statistics of appropriate measurements.  Likewise, the no-cloning theorem implies the impossibility of state measurement, since such a measurement would enable us to manufacture as many copies of the state as we desire. 

Approximate state determination and cloning are nevertheless possible.  The traditional approach of quantum detection theory has been to consider a quantum system whose unknown state belongs to a finite, known set, and to devise the measurement which yields the most information about the preparation, where the figure of merit is either the probability of a correct result or the mutual information\cite{Helstrom}.  Recently, however, Massar and Popescu\cite{Massar} and Derka {\em et al}\cite{Derka} have considered the problem of estimating a completely unknown quantum state, given $M$ independent realisations.  Analogous approaches have been used in the study of cloning.  Much of the work done here has focussed upon the issue of producing clones which resemble the original, as measured by the fidelity, as closely as possible\cite{Buzek,Hillery,Gisin,Bruss1,Bruss2}.  As with discrimination, attention has been given to the distinct problems of constructing cloning machines designed to optimally reproduce a finite number of states\cite{Hillery,Bruss1}, and universal cloning machines\cite{Buzek,Bruss1,Bruss2}, which are able to copy every quantum state with equal fidelity.
 
Non-orthogonal quantum states can, with some probability, be discriminated and cloned without approximation.  Ivanovic\cite{Ivanovic} showed that it is possible to discriminate {\em unambiguously}, that is, with zero error probability, between a pair of non-orthogonal states.  This measurement has subsequently been streamlined by Dieks\cite{Dieks} and Peres\cite{Peres}, and extended by Jaeger and Shimony\cite{JShimony} and by us\cite{Me1,Me2}.  More recently, Duan and Guo\cite{Duanguo1} have found that a pair of non-orthogonal states can be exactly cloned.  Both operations have interesting features in common.  For example, they are characterised by two possible outcomes: success or failure.  Unambiguous discrimination attempts may give inconclusive results, and the cloning operation may fail to copy the state.  In both cases, we can know with certainty whether or not the operation has been a success\cite{note1}.   

It is natural to enquire as to whether these similarities are the result of a deeper connection between these operations.  One particularly conspicuous common feature is that both operations transform the states in a manner which decreases their square-overlap.  In state discrimination, they are mapped onto orthogonal states, while in cloning, the square-overlap is itself squared.  Such operations may be said to perform {\em quantum state separation}.  The purpose of this Letter is two-fold.  We examine the properties of state separating operations for a pair of non-orthogonal states, and derive the maximum attainable value of the probability $P_{S}$ that two states can be separated by an arbitrary degree, assuming they have equal {\em a priori} probabilities.  We then discuss the relationship between state separation, unambiguous discrimination and cloning.  In particular, we show that the Ivanovic-Dieks-Peres bound\cite{Ivanovic,Dieks,Peres} on the probability of state-discrimination is a special case of our general bound on $P_{S}$.  We also explain how our earlier work\cite{Me1} on interpolation between unambiguous and optimal state-discrimination measurements can be understood in terms of state separation.  We then show that the Duan-Guo bound\cite{Duanguo1} on the probability of exactly cloning a pair of non-orthogonal states is also a particular case of the separation bound.   Furthermore, we find the more general least upper bound on the probability of producing $N$ copies of the state given $M$ initial ones, where $N{\ge}M$.  This result also follows directly from our bound on $P_{S}$.  Finally, we show that the state discrimination and $N$ from $M$ cloning bounds imply each other.  

Consider a quantum system prepared in one of the two  states $|{\psi}^{1}_{\pm}{\rangle}$.  We are not told which of the states the system is in, although we do know that it has equal probability of being in either.  We aim to transform the state into $|{\psi}^{2}_{\pm}{\rangle}$, where

\begin{equation}
|{\langle}{\psi}^{2}_{+}|{\psi}^{2}_{-}{\rangle}|^{2}{\le}|{\langle}{\psi}^{1}_{+}|{\psi}^{1}_{-}{\rangle}|^{2},
\end{equation}
that is, the operation decreases the square-overlap between the two possible states of the system, thus making them more distinct.  This operation cannot be successful all of the time, and there must be a probability of failure.  We wish to maximise the probability $P_{S}$ of this separation of the states being successfully carried out.   To analyse this problem, it is convenient to employ the Kraus representation of quantum operations\cite{Kraus}.  Here, each of the possible, distinguishable outcomes of an operation, which are labelled by the index $n$, is associated with a set of linear transformation operators ${\hat A}_{nk}$.  These form a resolution of the identity

\begin{equation}
\sum_{nk}{\hat A}^{\dagger}_{nk}{\hat A}_{nk}={\id}.
\end{equation}
The additional index $k$ allows for the possibility of different processes leading to the same outcome\cite{note2}.  Taking the system to be prepared with the initial density operator ${\hat \rho}$, the probability $P_{n}$ of the $n$th outcome is $\sum_{k}{\mathrm
Tr{\hat \rho}}{\hat A}^{\dagger}_{nk}{\hat A}_{nk}$.  Correspondingly, the density operator
is transformed according to ${\hat \rho}{\rightarrow}\sum_{k}{\hat A}_{nk}{\hat
\rho}{\hat A}^{\dagger}_{nk}/P_{n}$.

The state separation operation can have two possible outcomes, success or failure, which implies the respective transformation operators ${\hat A}_{Sk}$ and ${\hat A}_{Fk}$.  These operators act in the following way:

\begin{eqnarray}
{\hat A}_{Sk}|{\psi}^{1}_{\pm}{\rangle}&=&{\mu}_{k\pm}|{\psi}^{2}_{\pm}{\rangle},  \\ *
{\hat A}_{Fk}|{\psi}^{1}_{\pm}{\rangle}&=&{\nu}_{k\pm}|{\phi}_{k\pm}{\rangle},
\end{eqnarray}
where the ${\mu}_{k\pm}$ and ${\nu}_{k\pm}$ are complex coefficients and the $k$th failure operator transforms the $|{\psi}^{1}_{\pm}{\rangle}$ into some other normalised states $|{\phi}_{k\pm}{\rangle}$.  The resolution of the identity Eq. (2) becomes

\begin{equation}
\sum_{k}{\hat A}^{\dagger}_{Sk}{\hat A}_{Sk}+{\hat A}^{\dagger}_{Fk}{\hat A}_{Fk}={\id}.
\end{equation}
Consequently, we have $\sum_{k}|{\mu}_{k\pm}|^{2}+|{\nu}_{k\pm}|^{2}=1$.  We denote by $P_{S{\pm}}$ the conditional probability that the desired transformation takes place given the initial state $|{\psi}^{1}_{\pm}{\rangle}$, and find that  $P_{S{\pm}}=\sum_{k}|{\mu}_{k\pm}|^{2}$, implying that the total success probability is
\begin{equation}
P_{S}=\frac{P_{S+}+P_{S-}}{2}=\frac{1}{2}\sum_{k}|{\mu}_{k+}|^{2}+|{\mu}_{k-}|^{2}.
\end{equation}
This probability is bounded by the positivity of the operators ${\hat A}^{\dagger}_{Fk}{\hat A}_{Fk}$.  When combined wih the identity in Eq. (5), this implies that none of the eigenvalues of the operator $\sum_{k}{\hat A}^{\dagger}_{Sk}{\hat A}_{Sk}$ can exceed 1.  Consider any state $|{\psi}{\rangle}$ which is a superposition of $|{\psi}^{1}_{\pm}{\rangle}$: \begin{equation}
|{\psi}{\rangle}=N^{-1/2}\sum_{r=\pm}c_{r}|{\psi}^{1}_{r}{\rangle},
\end{equation}
where $\sum_{r}|c_{r}|^{2}=1$ and the normalisation factor $N$ is found to be $\sum_{r,r'}c^{*}_{r'}c_{r}{\langle}{\psi}^{1}_{r'}|{\psi}^{1}_{r}{\rangle}$.  The condition that ${\langle}{\psi}|\sum_{k}{\hat A}^{\dagger}_{Sk}{\hat A}_{Sk}|{\psi}{\rangle}{\le}1$ can be written as
\begin{equation}
\left( \begin{array}{rr}
    c^{*}_{+} &  c^{*}_{-} \\ \end{array} \right)
 \left( \begin{array}{cc}
    P_{S+} & Q{\beta}-{\alpha} \\ Q^{*}{\beta}^{*}-{\alpha}^{*} & P_{S-}
    \end{array}
\right) \left( \begin{array}{c}
   c_{+} \\  c_{-}
    \end{array}
\right){\le}1,
\end{equation} 
where $Q=\sum_{k}{\mu}^{*}_{k+}{\mu}_{k-}, {\alpha}={\langle}{\psi}^{1}_{+}|{\psi}^{1}_{-}{\rangle}$ and ${\beta}={\langle}{\psi}^{2}_{+}|{\psi}^{2}_{-}{\rangle}$.  Thus, we require the maximum eigenvalue of the Hermitian matrix in (8) to be no greater than 1, or, equivalently, that
\begin{equation}
(1-P_{S+})(1-P_{S-}){\ge}|{\alpha}-Q{\beta}|^{2}.
\end{equation}
This inequality can be used to obtain a bound on $P_{S}$ in the following way.  Firstly, we note that 
\begin{equation}
(1-P_{S})^{2}{\ge}(1-P_{S+})(1-P_{S-}),
\end{equation}
where the equality is satisfied only when $P_{S{\pm}}=P_{S}$.  Turning our attention  to the right-hand side of (9), we see that the triangle inequality gives
\begin{equation}
|{\alpha}-Q{\beta}|{\ge}|{\alpha}|-|Q||{\beta}|,
\end{equation}
and the equality here can only be attained when the phases of ${\alpha}$ and $Q{\beta}$ are the same.  The Cauchy-Schwarz inequality gives the following bound on $|Q|$:
\begin{equation}
|Q|{\le}(P_{S+}P_{S-})^{1/2}{\le}P_{S}.
\end{equation}
Here, the first equality is satisfied only when ${\mu}_{k-}$ is proportional to ${\mu}_{k+}$, and the second is equivalent to that in (10).  Together, these imply that ${\mu}_{k-}={\mu}_{k+}e^{i{\theta}}$, for some angle ${\theta}$, which is then also the phase of $Q$.  Combining this inequality with (11), and using $|{\alpha}|{\ge}|Q||{\beta}|$, we obtain
$|{\alpha}-Q{\beta}|{\ge}|{\alpha}|-P_{S}|{\beta}|$.  This equality can only be satisfied when the phases of the ovelaps between the initial states $|{\psi}^{1}_{\pm}{\rangle}$ and the (unnormalised) final states ${\mu}_{k{\pm}}|{\psi}^{2}_{\pm}{\rangle}$ are equal.  This,  when combined with (9) and (10), finally gives the separation bound  
\begin{equation}
P_{S}{\le}\frac{1-|{\langle}{\psi}^{1}_{+}|{\psi}^{1}_{-}{\rangle}|}{1-|{\langle}{\psi}^{2}_{+}|{\psi}^{2}_{-}{\rangle}|}.
\end{equation}
For this bound to be attained, both states must have equal conditional separation probabilities.  Also, the the phases of the overlaps between the initial and final states must be the same.  This can easily be satisfied as one of the states can be multiplied by an arbitrary phase factor without altering its physical meaning.

The maximum value of the separation probability has a natural composition property.  Consider three pairs of quantum states, $|{\psi}^{i}_{\pm}{\rangle}, i=1,2,3$, such that 
\begin{equation}
|{\langle}{\psi}^{3}_{+}|{\psi}^{3}_{-}{\rangle}|^{2}{\le}|{\langle}{\psi}^{2}_{+}|{\psi}^{2}_{-}{\rangle}|^{2}{\le}|{\langle}{\psi}^{1}_{+}|{\psi}^{1}_{-}{\rangle}|^{2}.
\end{equation}
and where the overlaps between all three sets of states have equal phases.  Then, denoting by $P^{12}_{S}, P^{23}_{S}$ and $P^{13}_{S}$ the maximum probabilities of carrying out the transformations $|{\psi}^{1}_{\pm}{\rangle}{\rightarrow}|{\psi}^{2}_{\pm}{\rangle}, |{\psi}^{2}_{\pm}{\rangle}{\rightarrow}|{\psi}^{3}_{\pm}{\rangle}$ and $|{\psi}^{1}_{\pm}{\rangle}{\rightarrow}|{\psi}^{3}_{\pm}{\rangle}$ respectively, we see that $P^{13}_{S}=P^{12}_{S}P^{23}_{S}$.  Thus, it is possible to attain the maximum separation probability by mapping the initial states directly onto the final ones, or through one or more sets of intermediate states.  

It is important to examine what happens when the separation does not occur, in particular whether or not one can make a further attempt.  If the attempt fails, then the initial density operator is transformed into
\begin{equation}
{\hat \rho}_{F\pm}=\frac{{\sum_{k}}{\hat A}_{Fk}|{\psi}^{1}_{\pm}{\rangle}{\langle}{\psi}^{1}_{\pm}|{\hat A}^{\dagger}_{Fk}}{{\sum_{k}}{\langle}{\psi}^{1}_{\pm}|{\hat A}^{\dagger}_{Fk}{\hat A}_{Fk}|{\psi}^{1}_{\pm}{\rangle}}.
\end{equation}
We will now show that if $P_{S}$ takes its maximum value, corresponding to saturation of the inequality (13), then the states ${\hat \rho}_{F\pm}$ are identical.  If the equality in (13) is satisfied, the operator $\sum_{k}{\hat A}^{\dagger}_{Sk}{\hat A}_{Sk}$ has an eigenvector lying in the subspace spanned by $|{\psi}^{1}_{\pm}{\rangle}$ with eigenvalue 1.  It follows from Eq. (5) that this eigenvector lies in the nullspace of $\sum_{k}{\hat A}^{\dagger}_{Fk}{\hat A}_{Fk}$.  Thus, there exists a state $|{\psi}{\rangle}$ of the form shown in Eq. (7) such that  $\sum_{k}{\langle}{\psi}|{\hat A}^{\dagger}_{Fk}{\hat A}_{Fk}|{\psi}{\rangle}=0$.  The positivity of the ${\hat A}^{\dagger}_{Fk}{\hat A}_{Fk}$ implies that the expectation value of each of these operators for the state  $|{\psi}{\rangle}$ must be zero, so every ${\hat A}_{Fk}$ annihilates $|{\psi}{\rangle}$.  Therefore, the expansion coefficients $c_{\pm}$ for the state $|{\psi}{\rangle}$ satisfy
\begin{equation}
c_{+}{\hat A}_{Fk}|{\psi}^{1}_{+}{\rangle}=-c_{-}{\hat A}_{Fk}|{\psi}^{1}_{-}{\rangle}.
\end{equation}
This expression, when applied to Eq. (15), immediately gives ${\hat \rho}_{F+}={\hat \rho}_{F-}$.  Thus, if an optimum state-separating operation fails, it is futile to make a further attempt to separate the states, since a failure will erase the bit of information describing the initial preparation. 
    
For the purposes of determining the general limit on $P_{S}$, we allowed for the possibility of there being multiple processes leading to either success or failure of the separation attempt.  We can simplify matters by considering an operation with just one process corresponding to each outcome.  A separation operator which attains the limit in (13) is
\begin{equation}
{\hat A}_{S}=\left(\frac{1-|{\langle}{\psi}^{1}_{+}|{\psi}^{1}_{-}{\rangle}|}{1-|{\langle}{\psi}^{2}_{+}|{\psi}^{2}_{-}{\rangle}|}\right)^{1/2}\sum_{r=\pm}\frac{|{\psi}^{2}_{r}{\rangle}{\langle}{\psi}^{1\perp}_{r}|}{{\langle}{\psi}^{1\perp}_{r}|{\psi}^{1}_{r}{\rangle}},
\end{equation}
where, of course, the phases of ${\langle}{\psi}^{1}_{+}|{\psi}^{1}_{-}{\rangle}$ and ${\langle}{\psi}^{2}_{+}|{\psi}^{2}_{-}{\rangle}$ are equal.  Here, we have introduced the states $|{\psi}^{1\perp}_{+}{\rangle}$ and $|{\psi}^{1\perp}_{-}{\rangle}$.  These are the states in the subspace spanned by the $|{\psi}^{1}_{\pm}{\rangle}$ which are orthogonal to $|{\psi}^{1}_{-}{\rangle}$ and $|{\psi}^{1}_{+}{\rangle}$ respectively.  The corresponding failure operator may be taken to have the form ${\hat A}_{F}={\hat U}({\id}-{\hat A}^{\dagger}_{S}{\hat A}_{S})^{1/2}$, where ${\hat U}$ may be any unitary operator.  

An interesting limit arises when $|{\psi}^{2}_{\pm}{\rangle}$ are orthogonal.  A von Neumann measurement would be able to distinguish perfectly between these states, and we would know, with absolute certainty, in which of the non-orthogonal states the system was prepared.  Given two non-orthogonal states $|{\psi}^{1}_{\pm}{\rangle}$ and setting  ${\langle}{\psi}^{2}_{+}|{\psi}^{2}_{-}{\rangle}=0$, the bound (13) on the separation probability is the limit on the probability of such an unambiguous discrimination attempt succeeding.  The maximum probability so obtained is equal to that derived by Ivanovic\cite{Ivanovic}, Dieks\cite{Dieks} and Peres\cite{Peres} in their studies of this specific problem, and is
\begin{equation}
P_{IDP}=1-|{\langle}{\psi}^{1}_{+}|{\psi}^{1}_{-}{\rangle}|. 
\end{equation} 
When the discrimination attempt fails, it will give an inconclusive result.  The IDP limit is not the absolute maximum of the discrimination probability, but is rather the maximum subject to the constraint that the measurement never gives incorrect results.   The absolute maximum probability of discriminating between two states $|{\psi}_{\pm}{\rangle}$ is instead given by the well-known Helstrom limit\cite{Helstrom}:
\begin{equation}
P_{H}=\frac{1}{2}\left(1+\sqrt{1-|{\langle}{\psi}_{+}|{\psi}_{-}{\rangle}|^{2}}\right),
\end{equation}
The Helstrom measurement does not give inconclusive results, but will incorrectly identify the state with probability $1-P_{H}$.      
Suppose that we map the states $|{\psi}^{1}_{\pm}{\rangle}$ onto $|{\psi}^{2}_{\pm}{\rangle}$, where $|{\psi}^{2}_{\pm}{\rangle}$ are not orthogonal but are more distinct than the initial states.  Let us consider then sending the output state to an optimal detector.  Such a detector will correctly identify the output state with probability $P_{H}$ in Eq. (19), with $|{\psi}_{\pm}{\rangle}=|{\psi}^{2}_{\pm}{\rangle}$.  The probability of an erroneous result, given that the separation takes place, is $1-P_{H}$.  Thus, the probability of correctly determining the state of the system is $P_{D}=P_{S}P_{H}$.  Likewise, the probability of obtaining an incorrect result $P_{E}=P_{S}(1-P_{H})$, and the probability of obtaining an inconclusive result $P_{I}=1-P_{S}$.  We have the sum rule
\begin{equation}
P_{S}+P_{E}+P_{I}=1.
\end{equation}
It is interesting to ask: what is the bound on the the error probability $P_{E}$ given a fixed value of the separation probability $P_{S}$, or equivalently, the probability $P_{I}$ of an inconclusive result?  Rearranging the separation bound in (13) to give an inequality for $|{\langle}{\psi}^{2}_{+}|{\psi}^{2}_{-}{\rangle}|$ in terms of $P_{S}$ and $|{\langle}{\psi}^{1}_{+}|{\psi}^{1}_{-}{\rangle}|$, which can be written in terms of $P_{IDP}$, we obtain
\begin{equation}
P_{E}{\ge}\frac{1}{2}\left(P_{S}-\sqrt{P^{2}_{S}-(P_{S}-P_{IDP})^{2}}\right).
\end{equation} 
When $P_{S}=1$, corresponding to no state separation, the minimum value of $P_{E}$ is $1-P_{H}$, which corresponds to the Helstrom measurement.  For $P_{S}=P_{IDP}$, we find that $P_{E}$ can be zero, which gives the IDP measurement.  This inequality is equivalent to that derived in \cite{Me1}, and corresponds to a family of measurements which optimally interpolates between the Helstrom and IDP limits.

A further application of state separation is the production of exact copies of a quantum system.  Suppose we have a quantum system in either of the states $|{\psi}_{\pm}{\rangle}$, again with equal {\em a priori} probabilities, and another system in the `blank' state $|{\chi}{\rangle}$.  A cloning operation will transform the product states $|{\psi}_{\pm}{\rangle}|{\chi}{\rangle}$ into $|{\psi}_{\pm}{\rangle}|{\psi}_{\pm}{\rangle}$.  Clearly, the square-overlap of the final states is the square of that of the initial states, and so is reduced by the operation.  Exact cloning is then seen to be a further example of state separation, and its success probability is bounded accordingly by (13).  Taking $|{\psi}^{1}_{\pm}{\rangle}=|{\psi}_{\pm}{\rangle}|{\chi}{\rangle}$ and $|{\psi}^{2}_{\pm}{\rangle}=|{\psi}_{\pm}{\rangle}|{\psi}_{\pm}{\rangle}$, we find that the separation bound in (13), interpreted as a bound on the probability of producing an exact copy of the initial state, gives the Duan-Guo limit\cite{Duanguo1}
\begin{equation}
P_{DG}=\frac{1}{1+|{\langle}{\psi}_{+}|{\psi}_{-}{\rangle}|}.
\end{equation}  

It is simple matter to use the bound on $P_{S}$ to derive the least upper bound on producing $N$ copies of $|{\psi}_{\pm}{\rangle}$  from $M$ initial copies where $N{\ge}M$.  We take the initial states $|{\psi}^{1}_{\pm}{\rangle}$ to be the products $|{\psi}_{\pm}{\rangle}_{1}{\ldots}|{\psi}_{\pm}{\rangle}_{M}|{\chi}{\rangle}$, where the blank state $|{\chi}{\rangle}$ is composed of $N-M$ subsystems.  The final states $|{\psi}^{2}_{\pm}{\rangle}$ are  $|{\psi}_{\pm}{\rangle}_{1}{\ldots}|{\psi}_{\pm}{\rangle}_{N}$, the $N$ copies of $|{\psi}_{\pm}{\rangle}$.  It immediately follows from (13) that the maximum probability of producing $N$ copies given a smaller number $M$ of initial copies of the states $|{\psi}_{\pm}{\rangle}$ is
\begin{equation}
P_{MN}=\frac{1-|{\langle}{\psi}_{+}|{\psi}_{-}{\rangle}|^{M}}{1-|{\langle}{\psi}_{+}|{\psi}_{-}{\rangle}|^{N}},
\end{equation} 

Let us finally examine the relationship between of the general cloning bound $P_{MN}$ and the Ivanovic-Dieks-Peres bound $P_{IDP}$, by showing that they can be derived from each other.  To derive the latter from the former, suppose that we possess a single system prepared in one of the states $|{\psi}_{\pm}{\rangle}$.  The maximum probability that we can make $N$ copies of the state is given by $P_{1N}$.  We can see from Eq. (23) that as $N{\rightarrow}{\infty}$, $P_{1N}{\rightarrow}P_{IDP}$ from above.  In this limit, the state could be inferred through the statistics of appropriate measurements on the copies, so we have shown how Eq. (23) implies that the states can be discriminated unambiguously with probability $P_{IDP}$.  Duan and Guo have also considered state-identification for arbitrary numbers of states to be equivalent to the production of an infinite number of copies\cite{Duanguo2}.   Consistency with the cloning bound implies that no greater value than $P_{IDP}$ can be attained.  If state discrimination could be accomplished with probability higher than $P_{IDP}$, then we could, with the same probability, make an arbitrarily large number of copies of the state given one initial realisation.  If this probability was greater than $P_{IDP}$, it would also exceed $P_{1N}$ for sufficiently large $N$.   The discrimination limit $P_{IDP}$ and the infinite cloning bound $P_{1{\infty}}$ (or, for that matter, the maximum discrimination probability with $M$ copies of the state, and $P_{M{\infty}}$), are equal due to the equivalence of these operations. 

We can also prove that the discrimination limit $P_{IDP}$ implies that $P_{MN}$ is an upper bound on the general cloning probability.  Suppose that we have $M$ quantum systems all prepared in one of the states $|{\psi}_{\pm}{\rangle}$.  If $P_{IDP}$ is the maximum probability that we can discriminate between two quantum states, then it is impossible to discriminate between the $M$-particle products with probability greater than $P_{M\infty}$.  Thus, we ought to be unable to improve upon this bound by first attempting an $N$ from $M$ cloning operation, with $N{\ge}M$, followed by an attempt to discriminate between the $N$-particle products, which can be accomplished with probability no greater than $P_{N\infty}$.  The cloning probability, which we shall write as $P_{C}$,  must be constrained by the fact this compound operation cannot be accomplished with probability greater than $P_{M\infty}$.  Thus, $P_{M\infty}{\ge}P_{C}P_{N\infty}$, so $P_{C}$ cannot be greater than $P_{MN}$ in (23).  

We have shown that it is possible, with non-zero probability, to map a pair of quantum states onto another pair with a lower square-overlap.  If the states have equal {\em a priori} probabilities, then the maximum attainable value of the separation probability $P_{S}$ is given by the inequality in (13).  This limit can account for the established bounds on the probabilities of performing unambiguous discrimination and exact cloning of two states, and also directly yields the maximum probability $P_{NM}$ of producing $N$ copies given $M$ initial ones, with $N>M$.  The fact that the latter bound is strongly related to that on the discrimination probability should not surprise us.  Neither the no-cloning theorem nor the impossibility of infallible state discrimination can be consistent without the other.  Here, we have shown how the relationship between  $P_{MN}$ and $P_{IDP}$ expresses their mutual consistency.  This may be seen to complement the recent results of Gisin and Massar\cite{Gisin} and Bru{\ss} {\em et al}\cite{Bruss2} which relate the maximum fidelity of a universal, {\em approximate} cloning machine to limits on state estimation.  

It is possible that all of the  qualitatively different approaches to state discrimination correspond to a particular type of cloning machine, with the bounds on their appropriate figures of merit being related through consistency.  This perhaps suggests that a more general, quantitative relationship between cloning and discrimination limits can be found; one that transcends the  many particular strategies and deepens our understanding of both types of operation in general.      

\section*{Acknowledgements}
We gratefully acknowledge financial support by the UK Engineering and Physical Sciences Research Council (EPSRC).


\begin{references}
\bibitem{note1}
Also, when extended to the case of multiple quantum states, it has been found that both types of operation are possible only when the state belongs to a known linearly-independent set (see \cite{Me2,Duanguo2}).
\bibitem{note2}
The range of $k$ can be taken to be the same for all outcomes, since we may allow  some of the ${\hat A}_{nk}$ to be zero.
\end{references}
\end{document}